# Nonlinear Optical Spectroscopy of Excited States in Polyfluorene

M. Tong, C. -X. Sheng and Z. V. Vardeny*

*Department of Physics, University of Utah, Salt Lake City, Utah 84112*

We used a variety of nonlinear optical (NLO) spectroscopies to study the singlet excited states order, and primary photoexcitations in polyfluorene; an important blue emitting π-conjugated polymer. The polarized NLO spectroscopies include ultrafast pump-probe photomodulation in a broad range of 0.2 to 2.6 eV, two-photon absorption in the range of 3.2 - 4.2 eV, and electroabsorption covering the spectral range of 2.8 - 5.3 eV. For completeness we also measured the linear absorption and photoluminescence spectra. We found that the primary photoexcitations in polyfluorene are singlet excitons with ~ 100 ps lifetime that have a characteristic photomodulation spectrum comprising of two photoinduced absorption (PA) bands, $PA_1$ at 0.55 eV and $PA_2$ at 1.65 eV, respectively, and a strong stimulated emission band peaked at ~ 2.5 eV. The two-photon absorption and electroabsorption spectra identify the exciton PA bands with optical transitions between the lowest lying odd symmetry $1B_u$ exciton at 3.1 eV into two strongly coupled even symmetry states, namely $mA_g$ at 3.7 eV and $kA_g$ at 4.7 eV, respectively. The excited states manifold also contains a strongly coupled odd symmetry exciton, $nB_u$ at 4.1 eV. A polarization memory of ~ 2.2 typical to π-conjugated polymer chains characterizes all three NLO spectra reflecting the highly anisotropic third-order NLO coefficient $\chi^{(3)}$ of the polymer chains. The four essential excited states, namely the $1B_u$, $mA_g$, $nB_u$ and $kA_g$ were used to satisfactorily fit all three nonlinear optical spectra using the summation over states model. The combination of the three NLO spectra and the model fit conclusively show that the band model typical to inorganic semiconductors cannot describe the PFO polymer. On the contrary, a strongly bound exciton with intrachain binding energy of ~ 0.9 eV dominates the linear and NLO spectra of this polymer.

* To whom correspondence should be addressed; e-mail: val@physics.utah.edu

1. **Introduction**

Π-conjugated polymers have emerged as a promising class of organic semiconductors due to their low processing cost and the broad range over which their optical and electronic properties may be chemically tuned [1]. Several device applications have been commercialized such as bright organic light emitting diodes (OLED) with a variety of colors ranging from red to blue, and organic field effect transistors; other applications have been investigated such as white OLED, large area photodetectors, organic photovoltaic cells, and current injected lasers [2, 3]. However in spite of tremendous research of more than a decade, the photophysics of π-conjugated polymers is still hotly debated [4]. For example, it is still unclear whether the absorption spectrum of π-conjugated polymers is due to interband transition (namely valence to conduction band), where photon absorption results in free carriers as in inorganic semiconductors [5, 6]; or it is excitonic in nature as in molecular solids resulting in singlet excitons as the primary photoexcitations [7]. It is generally accepted that photoluminescence (PL) is due to intrachain excitons rather than interband transition of photogenerated electrons (e) and holes (h). However the formation of secondary interchain species in π-conjugated polymer films that are nonradiative charge transfer excitons cannot be ruled out [8]. One of the reasons for this confusion is that many π-conjugated polymers show a branching ratio of photogenerated excitons and charge polarons following photon absorption [9, 10]. This branching ratio has been invoked to be due to interchain interaction between adjacent chains [11], which should decrease with larger interchain separation when bulky side groups are involved in the polymer structure.

Polyfluorene is an attractive material for display applications due to efficient *blue emission* [12] and relatively large hole mobility with trap free transport [13]. Poly(9,9-dioctylfluorene) [PFO] (shown in Fig. 1 inset) exhibits a complex morphological behavior that had interesting implications for its photophysical properties [14, 15]. It was previously shown that the structural versatility of PFO can be exploited in manipulating the sample electronic and optical properties [14, 16]. When changing a pristine sample with glassy structure dubbed the α phase, into a film with more superior order dubbed β phase, then hole transport increases [16], laser action occurs at reduced excitation intensities, and spectral narrowing is obtained at different wavelengths [17]. However, even in the disordered, α phase PFO shows relatively high degree of planarity, which together with the bulky side group should provide a clean case for studying intrachain photoexcitations and characteristic excited states with only a small contribution due to interchain interaction. Nevertheless several early studies of ultrafast photoexcitation dynamics in PFO have led to confusing results. In one study of oriented PFO [18], three types of photoexcitations were invoked including hot carriers, excitons, and charge polarons. In two other studies [19, 20], both excitons and bound polaron pairs were shown to simultaneously coexist. Under these circumstances it is difficult to decide whether PFO excited states are band like or excitonic in nature. Some of the reasons for this confusion is the relatively narrow spectral range in which the PFO photoexcitations were previously probed, the high excitation intensity used, as well as the lack of other complementary optical measurement techniques to fully understand the nature of the excited states in this polymer. In the quest to understand the singlet manifold of PFO more recent studies have focused on three beam excitation [21-23], where the role of even parity states was emphasized. In these studies it was realized that when excited deeper into the singlet manifold charge species are primarily photogenerated; however most of them geminately recombine within few ps.

In the present work we employ a variety of nonlinear optic (NLO) spectroscopies to elucidate the photoexcitations and important excited states in the singlet manifold of PFO. The NLO spectroscopies include ultrafast pump-probe photomodulation (PM), electroabsorption (EA), and two-photon absorption (TPA); we complete the optical investigation by comparing the NLO spectra with linear optical measurements that include absorption and PL spectra. TPA spectroscopy is sensitive to excited states with even, $A_g$ symmetry, and thus is complementary to linear absorption that probes excited states with odd or $B_u$ symmetry. EA spectroscopy is sensitive to excited states of both odd, $B_u$ and $A_g$ symmetry. The TPA spectrum was measured before in another PFO derivative in solution [24]; whereas the EA spectrum was measured in PFO film that underwent several phase transitions [14], but the spectra were analyzed in terms of first and second derivatives of the absorption spectrum without underlying the role of specific electronic excited states of the polymer. The application of various NLO and linear spectroscopies, such as the transient PM, TPA and EA, together with the linear absorption and PL spectra, on the *same* PFO film has provided a more complete picture of the electronic states in PFO compared to previous works.

The NLO spectra were analyzed in terms of several essential excited states [25] that are strongly coupled to each other and to the ground state, namely $1B_u$, $mA_g$, $nB_u$, and $kA_g$ excitonic states, using the summation over states method [26]; and a consistent and elegant picture for PFO has emerged. We found that the primary photoexcitations in PFO are singlet excitons with two photoinduced absorption (PA) bands in the mid- and near-ir spectral ranges, respectively; and a strong stimulated emission (SE) band in the visible range. The PA bands can be understood [27] as resulting from optical transitions in the singlet manifold, namely from $1B_u$ to $mA_g$ and $1B_u$ to $kA_g$, respectively; and SE from the $1B_u$ to the ground (or $1A_g$) state. We conclude that PFO excited state manifold is composed of several excitonic states with odd ($B_u$) and even ($A_g$) symmetry, rather than a continuum band that characterizes regular semiconductors. From our experimental results we infer that the lowest intrachain excited electronic state in PFO is a strongly bound exciton ($1B_u$), having an intrachain binding energy of ~ 0.9 eV.

2. **Experimental**

The transient PM spectrum of the PFO films was studied using the polarized pump-and-probe correlation technique with ~ 150 fs time resolution. For this study we utilized two fs Ti:sapphire laser systems with low and high repetition rates, having high and low pulse energies, respectively [7]. These two laser systems separately covered the mid- and near-ir spectral ranges; the high repetition laser was used in the mid-ir range [20], whereas the low repetition laser system was used in the near-ir and visible spectral range [28]. The pump pulses were kept at 3.2 eV for both laser systems. The PM spectra obtained using the two laser systems were then normalized to each other in the visible range (2 eV) providing a very broad probe spectral range from 0.15 to 2.6 eV, with only small spectral gaps.

The low repetition rate high-energy laser system was a homemade Ti:sapphire regenerative amplifier that provides pulses of 120 fs duration at photon energies of 1.55 eV, with 400 µJ energy per pulse at a repetition rate of 1 kHz. The second harmonic of the fundamental pulses at 3.2 eV was used as the pump beam. The probe beam was a white light supercontinuum within the spectral range from 1.2 to 2.7 eV, which was generated using a portion of the Ti:sapphire amplifier output in a 1 mm thick sapphire plate. An overall time resolution of ~ 150 fs in the pump-probe measurements was achieved by adjusting the stage to compensate the spectral chirp as measured by time-resolved TPA (see below). The probe beam polarization was set to be either parallel or perpendicular to the pump beam polarization. To improve the signal-to-noise ratio, the pump beam was synchronously modulated by a mechanical chopper at exactly half the repetition rate of the Ti:sapphire laser system (~ 500 Hz). The probe beam was mechanically delayed with respect to the pump beam using a computerized translation stage in the time interval, $t$ up to 200 ps. The delay line that corresponds to $t = 0$ was set by a sum frequency cross correlation trace of the pump and probe pulses in a NLO crystal.

Since some of the photoexcitation dynamics depend on the excitation density, care was taken in the experimental design to minimize distortion of the measured pump-probe response by spatial inhomogeneity of the photoexcitation distribution. The pump beam was focused onto the sample to a 1 mm diameter round spot, whereas the probe beam was focused using an achromatic lens onto a 0.4 mm diameter spot in the center of the pump illuminated spot. To ensure the reproducibility of the alignment, the spatial overlap of the pump-probe beams was set using a telescopic microscope. The wavelength resolution of this system was about 4 nm using a 1/8-meter monochromator with a 0.6 mm exit slit, which was placed in the probe beam after it passed through the sample. The transient spectrum of the photoinduced change ($\Delta T$) in the sample transmission ($T$) was obtained using a phase-sensitive technique. Following pump photon absorption the probe beam experience photoinduced absorption (PA), which is represented in the spectra as negative differential transmittance, $-\Delta T/T$. Since the pump and probe beams are linearly polarized we could measure $\Delta T_{pa}$ ($\Delta T_{pe}$), where the pump and probe polarization are parallel (perpendicular) to each other [28]. Pump induced stimulated emission (SE) and photobleaching (PB) of the optical absorption in the ground state with $\Delta T > 0$ were also measured [27]. In the small signal limit, $\Delta T(t)$ is expected to be proportional to the photoexcitation density, $N(t)$, which for an optically thick film is given by the relation: $\Delta T/T = N\sigma/\alpha_L$, where $\sigma$ is the photoexcitation optical cross-section, and $\alpha_L$ is the absorption coefficient at the pump laser excitation wavelength. To correct the pump-probe signal for intensity fluctuations in the supercontinuum at the selected probe wavelength, the probe signal was normalized by a reference signal at each delay time (this technique was dubbed as "A-B" [29]) with a significant improvement in the measured signal sensitivity of up to $\Delta T/T = 10^{-4}$ that corresponds to a photoexcitation density of about $10^{17}$ cm$^{-3}$.

The high repetition rate low energy laser system was an optical parametric oscillator (OPAL from Spectra Physics) that was pumped by a 100 fs Ti:sapphire laser oscillator (Tsunami, Spectra Physics) at a repetition rate of about 80 MHz [28]. The pump beam was extracted from the laser oscillator and was frequency doubled to 3.2 eV. The probe beam was extracted from the signal and idler beams of the optical parametric oscillator in the spectral ranges 0.56–0.68 eV and 0.94–1.02 eV, respectively, with about 150 fs time resolution. We extended the spectral range deeper into the ir using difference frequency set up in a NLO crystal. When using the signal and idler beams, and changing the central frequency of the OPO to 1.55 μm we could extend the spectral range between 0.15 – 1.05 eV with few local gaps in the spectrum [7]. This laser system provides low-intensity measurements, where the pump intensity ranges from 0.1 to 30 mJ/cm$^2$ per pulse, with $\Delta T/T$ resolution $\approx 10^{-6}$ that corresponds to photoexcitation density of $10^{15}$ cm$^{-3}$.

PFO powder was purchased from ADS (in Canada) and was used to make films either by drop cast or spin cast on sapphire substrates, without further purification. For the pump-probe studies all measurements were carried out at room temperature in a cryostat that provided a dynamical vacuum of 100 μTorr for preventing the polymer film degradation due to the strong laser illumination at ambient conditions [30]. Pump-probe signals were measured over a range of pump intensities to ensure linearity of $\Delta T/T$ response with respect to the initial photoexcitation density; we thus work at intensities below 300 µJ/cm$^2$ per pulse, with typical intensities of about 30 µJ/cm$^2$ to prevent signal saturation.

The TPA spectrum was measured using the polarized pump-probe correlation technique with the low repetition rate high-energy laser system at time delay $t = 0$. The linearly polarized pump beam was set at 1.55 eV, below the polymer absorption band; whereas the probe beam from the white light supercontinuum covered the spectral range from 1.6 to 2.6 eV. The temporal and spatial overlap between the pump and probe beams on the sample film leads to a photoinduced absorption (PA) signal that peaks at $t = 0$. We interpret it here as due to TPA of one pump photon with one probe photon.

For the EA measurements we used a PFO film spin cast on a substrate with patterned metallic electrodes [26]. The EA substrate consisted of two interdigitated sets of a few hundred gold electrodes 30 µm wide patterned on a sapphire disk 1 mm thick. The sample was placed in a cryostat for low temperature measurements. An electric field was generated in the sample by applying a potential *V* to the electrodes. An applied potential $V = 600$ Volts (typical for our experiments) resulted in field strength $F = 2\times10^5$ V/cm. We varied *V* on the electrodes using a sinusoidal signal generator at f = 1 kHz, and a simple transformer to achieve high voltages. For probing the EA spectrum we used an incandescent light source from a Xe lamp, with broadband visible and ultraviolet spectral range from 2.6 to ~ 5.3 eV. The light beam was dispersed through a monochromator, focused on the sample, and detected by silicon photodiodes. The modulation, $\Delta T$ of the transmission, *T* is expected to be the same for positive and negative *V*'s, since the polymer chains are not preferentially aligned with respect to the electrodes. Thus, we measured $\Delta T$ using a lock-in amplifier set to twice the frequency (2*f*) of the applied field [26]. We verified that no EA signal was observed at *f* or at 3*f*. $\Delta T$ and *T* spectra were measured separately using a homemade spectrometer that consisted of a ¼ meter monochromator equipped with several gratings and solid-state detectors such as Si and Si-UV enhanced diodes to span the EA in the broadest spectral range. The EA spectrum was obtained from the ratio $\Delta T/T$, which was measured at various applied voltages and polarizations of the probe light respect to the direction of the applied field.

## 2. Results and discussion

*(i) Absorption and PL spectra*

The PFO polymer repeat unit is shown in Fig. 1 inset. The polymer chain is highly planar even in the disordered α phase, since the two repeat benzene rings are tied together with two large adjacent side groups. The room temperature absorption and PL spectra are shown together in Fig. 1 for ease of comparison. The main absorption band peaks at 3.2 eV with an onset at ~ 2.95 eV; however it is featureless and thus gives the impression of a smooth absorption spectrum, similar to that typical of inorganic semiconductors. The PL spectrum has a relatively small red shift respect to the absorption band, and exhibits a clear vibronic structure with peaks at about 2.90 (0-0), 2.72 (0-1), and 2.55 (0-2) eV, respectively; a fourth phonon side band (0-3) may be seen at ~ 2.37 eV. The vibrational series in the PL spectrum shows that the polymer possesses strong electron-phonon coupling to a strongly coupled vibration (identified as the C=C stretching), which, however is not clearly seen in the absorption spectrum. The reason for the apparent dissimilarity between the PL and absorption spectra is the existence of a broad distribution of the polymer conjugation length (CL), where the characteristic optical energy gap of the chains depends inversely on the CL. Whereas the absorption process occurs to all chains, the PL is preferentially emitted from the longest chains in the film having the smallest optical gap; and thus the phonon side bands seen in the PL spectrum are mainly related to these longer chains. We therefore conclude that the smooth absorption spectrum has little to do with interband transition, but in fact is an inhomogeneous broadened version of delocalized π- π* transitions involving optical transitions from the ground state ($1A_g$) to the first odd-parity exciton ($1B_u$) [31]. It is interesting to note that the absorption spectrum also contains a small feature at ~ 4.1 eV; this feature is enhanced, and thus seen more clearly in the EA spectrum (see below).

*(ii) Electroabsorption spectroscopy*

To elucidate the nature of the excited states responsible for the broad optical absorption band in PFO we have applied the EA spectroscopy. EA has provided a sensitive tool for studying the band structure of inorganic semiconductors [32], as well as their organic counterparts [33-35]. Transitions at singularities of the joint density of states respond particularly sensitively to an external field, and are therefore lifted from the broad background of the absorption continuum. The EA sensitivity decreases, however in more confined electronic materials, where electric fields of the order of 100 kV/cm are too small of a perturbation to cause sizable changes in the optical spectra. As states become more extended by inter-molecular coupling they respond more sensitively to an intermediately strong electric field, $F$ since the potential variation across such states cannot be ignored compared to the separation of energy levels. EA thus may selectively probe extended states and thus is particularly effective for organic semiconductors, which traditionally are dominated by excitonic absorption. One of the most notable examples of the application of EA spectroscopy to organic semiconductors is polydiacetylene, in which EA spectroscopy was able to separate absorption bands of quasi-1D excitons from that of the continuum band [36]. The confined excitons were shown to exhibit a quadratic Stark effect, where the EA signal scales with $F^2$ and the EA spectrum is proportional to the derivative of the absorption respect to the photon energy. In contrast, the EA of the continuum band scales with $F^{1/3}$ and shows Frank-Keldish (FK) type oscillation in energy. The separation of the EA contribution of excitons and continuum band was then used to obtain the exciton binding energy in polydiacetylene, which was found to be ~ 0.5 eV [36].

Fig. 2 shows the EA spectrum of a PFO film on sapphire substrate up to 5.3 eV, at field value $F$ of $10^5$ volt/cm. The EA spectrum was measured at 80 K to decrease the thermal effect contribution, which also has a $F^2$ dependence similar to that of the EA itself. There are five main spectral features in the EA spectrum: two derivative-like features with zero crossing at 3.1 (assigned as $1B_u$) and 4.1 eV (assigned as $nB_u$), respectively; two well-resolved phonon sidebands related to the $1B_u$ derivative feature at 3.2 and 3.4 eV, respectively; and two induced absorption bands at 3.7 eV (assigned as $mA_g$) and 4.5 eV (assigned as $kA_g$), respectively. No Frank Keldish type oscillation related to the onset of the interband transition [36] is seen in the EA spectrum here. In fact, the two main derivative-like features are due to Stark shift characteristic of excitons [34]. We thus conclude that the PFO excited states are better described in terms of excitons, rather than in the language of band-to-band transition typical of inorganic semiconductors. We also measured the polarization dependence of the EA spectrum. We found that the EA spectrum parallel to the direction of the applied field is about 2.5 larger than that perpendicular to the field; but otherwise the two spectra are very similar.

We interpret the EA spectrum as following [26, 35]: The first derivative like features at 3.1 eV and 4.1 eV are due to a Stark shift of the lowest lying exciton, $1B_u$ and the most strongly coupled exciton, $nB_u$. The second derivative like feature at energies just above $E(1B_u)$ is due to Stark shift of the $1B_u$ - related phonon side bands. These features are more easily observed in EA than in the linear absorption spectrum because of the strong dependence of the exciton polarizability on the CL in the polymer chains. The polarizability was shown [37] to increase as $(CL)^n$, where n ~ 6, and thus the EA spectrum preferentially focuses on long CL, similar to the case of the PL spectrum discussed above. The EA induced absorption feature at 3.7 eV does not have any corresponding spectral feature in the linear absorption spectrum. We therefore conclude that this feature in the EA spectrum involves a strongly coupled $A_g$ state, dubbed $mA_g$ [35]. Such a state would not normally show up in the linear absorption spectrum since the optical transition from the ground state $(1A_g)$ to $mA_g$ is forbidden. The presence of this band in the EA spectrum can be explained by the external electric field effect on the film, which breaks the symmetry resulting in the transfer of oscillator strength from the allowed $1A_g$ to $1B_u$ transition to the forbidden $1A_g$ to $mA_g$ transition [25]. The same applies for the EA feature at ~ 4.5 eV assigned as $kA_g$ [27]; this is another strongly coupled $A_g$ state that is further away from the $1B_u$, which, however may be strongly coupled to the $nB_u$ in the spectrum [38].

Electroabsorption is a third-order NLO effect and thus can be described by the third order optical susceptibility, namely $\chi^{(3)}(-\omega; \omega, 0, 0)$ [25, 35]:

$$-\Delta T / T = \frac{4\pi\omega}{nc} \text{Im}[\chi^{(3)}(-\omega;\omega,0,0)] F^2 d \quad (1)$$

where *d* is the film thickness, *n* is the refractive index, c is the speed of light, and ω is the optical frequency; the field modulation f << ω, and this explains the zero frequency for $\chi^{(3)}$ in Eq. (1). The relation between the EA and $\chi^{(3)}$ in Eq. (1) shows that the polarization dependence of the EA spectrum is in fact related to the anisotropy of $\chi^{(3)}$ for the PFO film. The obtained polarization of ~2.5 is thus not surprising since the PFO polymer chains are quite anisotropic, with NLO coefficient that is stronger along the polymer chains. We thus expect similar anisotropy to hold also for the other NLO measurements described here, which can be also described by $\chi^{(3)}$.

For obtaining more quantitative information about the main exciton states in PFO we fitted the EA spectrum using a model calculation. For this fit we calculated $\chi^{(3)}$ using the summation over states (SOS) model, originally proposed by Orr and Ward [39], and further developed for π-conjugated polymers by Mazumdar at al. [25, 35], and implemented in a variety of conducting polymers by Liess et al. [26]. In this model $\chi^{(3)}$ is a summation of 16 terms each containing a denominator that is proportional to multiplication of several dipole moment transitions, such as from the ground state to $1B_u$ and $1B_u$ to $mA_g$; and a nominator that is resonance at energies related to the main essential states [see Eqs. 3-12 in ref. 26]. Apart from the $kA_g$ state [38], and according to the "essential states picture" in π-conjugated polymers advanced by Mazumdar and colleagues [25, 35], there are four essential states that contribute substantially to the main EA spectral features in these compounds; these are $1A_g$, $1B_u$, $mA_g$, and $nB_u$. The essential states energies, and dipole moment transitions were taken as free parameters in the fit (see Table I). In the fitting we also took into account the main phonon side bands, with phonon frequency ~ 185 meV for to the most strongly coupled intrachain vibration (the C=C stretching mode); as well as an asymmetric CL distribution function [26]. The phonon side bands and the CL distribution were shown to be very important in fitting the EA spectra of many π-conjugated polymers [26]. The solid line in Fig. 2 shows the best fit to the EA spectrum as obtained using the parameters given in Table I; the agreement between the model calculation and experimental spectra is very good. From Table I we get the energies of the most strongly coupled excitons in PFO as following: $E(1B_u) = 3.1$ eV; $E(mA_g) = 3.7$ eV; and $E(nB_u) = 4.0$ eV. Furthermore from the EA spectrum at energies above $E(nB_u)$ we approximate $E(kA_g) = 4.7$ eV [38]. These essential states play an important role in the other two NLO spectroscopies, namely the TPA and transient PM spectra as discussed below.

From the magnitude of the EA signal compared to the first derivative of the absorption spectrum, an estimate of the difference in polarizability, $\Delta p$ between the ground and the first excited state, namely the $1B_u$ may be obtained [14, 26]. This requires a careful fit to the absorption spectrum using the same parameters that were used to fit the EA spectrum and a different CL distribution function [26]. Instead we estimate $\Delta p$ from a simple comparison of the PFO EA spectrum to that of a various polymers where $\Delta p$ is known [26]. We obtained a polarizability difference between $1A_g$ and $1B_u$ states, $\Delta p \sim 5000$ Å$^3$ in fair agreement with a previous estimate [14].

It is known that the continuum band in π-conjugated polymers is very close to $E(nB_u)$ [35]; it is thus tempting to identify the transition from the ground state into the continuum band from the linear absorption spectrum. Based on the EA spectrum at ~ 4 eV and its analysis in term of $E(nB_u)$, we then tentatively assign the absorption band II in the linear absorption spectrum (Fig. 1) at ~ 4.1 eV as the optical transition from the $1A_g$ to $nB_u$ (or continuum) state. We estimate the absorption strengths ratio of bands I and II in the linear absorption spectrum (Fig. 1) to be ~ 100:1. It is well established that in semiconductors the interband transition strength substantially decreases relative to the exciton transition strength for large exciton binding energy, $E_b$ [35, 36]. The large ratio obtained between bands I and II in the absorption spectrum thus indicates that $E_b$ in PFO is relatively large, of the order of 0.5 eV. Actually we may estimate $E_b$ of the lowest lying singlet exciton in PFO from the relation $E(nB_u) - E(1B_u)$. From the values given in Table I we get a large binding energy, $E_b \approx 0.9$ eV. This large $E_b$ is not unique in the class of π-conjugated polymers. It is similar to $E_b$ extracted for MEH-PPV ($E_b \approx 0.8$ eV) when using the corresponding EA spectrum [26]. The large value for $E_b$ shows that electron-hole interaction, and electron correlation [15] are relatively large in PFO, and do not permit to describe this polymer in terms used by the semiconductor band model [4].

*(iii) Two-photon-absorption spectroscopy*

In π-conjugated polymers the optical transitions between the ground state $1A_g$ and the $B_u$ excitonic states are allowed; in particular the transition between $1A_g$ to $1B_u$ dominates the absorption spectrum [40, 41]. On the contrary, the optical transitions between $1A_g$ to other states with $A_g$ symmetry are forbidden; however, these optical transitions are allowed in two-photon-absorption (TPA) [25]. Therefore TPA spectroscopy has been used in the class of π-conjugated polymers to get information about the $A_g$ energies in these materials [42, 43]. This information is important since transitions of photogenerated $1B_u$ excitons to $A_g$ states are dipole allowed, and thus dominate the PA spectrum of excitons in π-conjugated polymers [44]. In addition, it was also found [45] that the resonant Raman scattering dispersion known to exist in π-conjugated polymers, surprisingly depend on the lowest lying $A_g$ states, rather than the $B_u$ states; and thus $E(A_g)$ are worthy to determine in this class of materials.

Usually the TPA spectrum has been measured in π-conjugated polymers either directly by techniques such as optical absorption at high excitation intensity [46, 47], and Z-scan [48, 49]; or indirectly by measuring the fluorescence emission following TPA at high intensity; a technique dubbed two-photon-fluorescence [50]. Typically two TPA bands are observed in π-conjugated polymers, namely $mA_g$ and $kA_g$, where $E(mA_g) < E(kA_g)$ [42]. It is noteworthy mentioning that in nonluminescent π-conjugated polymers, such as t-$(CH)_x$ [46] and polydiacetylene [47], it was measured that $E(2A_g) < E(1B_u)$. In this case the photogenerated $1B_u$ excitons quickly undergo internal conversion into the lowest energy exciton that is the $2A_g$; consequently the transient PL emission is very fast, of the order of 200 fs. However it has been found [42] that the $2A_g$ state is not easy to detect by NLO techniques since it is not strongly coupled to any $B_u$ states, and thus is very weak in both TPA and EA spectra.

In the present work we have chosen to measure the TPA spectrum using the pump and probe technique at t = 0. The linearly polarized pump beam from the low repetition high power laser system was set at 1.55 eV, below the polymer absorption band; whereas the probe beam from the white light supercontinuum with polarization either parallel or perpendicular to that of the pump beam spreads the spectral range from 1.6 to 2.6 eV, thus covering the TPA photon energy from 3.15 to 4.15 eV. If only linear absorption is considered, then the pump beam alone is unable to generate photoexcitations above the gap, since its photon energy is much smaller than the optical gap of PFO at ~ 3.1 eV. However the temporal and spatial overlap between the pump and probe beams leads to a photoinduced absorption (PA) signal that peaks at t = 0. As seen in Fig. 3 this PA has a temporal profile similar to the cross-correlation function of the pump and probe pulses, which we interpret here as due to TPA of one pump photon with one probe photon. The long temporal tail seen in Fig. 3 at times longer than that of the cross correlation function is caused by PA of photoexcitations that are generated due to the TPA of the pump pulses alone [22]. This tail was subtracted out from the transient response at t = 0 to give the TPA spectrum clear from the PA due to the TPA-related photoexcitations. Otherwise the TPA related to the pump pulses does not directly influence the transmission of the probe pulses. Two separate TPA spectra were obtained; where the probe beam polarization set either parallel, or perpendicular to the pump beam polarization.

Figure 4(a) shows the TPA spectrum in PFO film up to 4.4 eV compared with the linear absorption spectrum. The TPA shows a relatively broad band peaked at 3.7 eV, which has comparable width to that of the linear absorption band. We interpret the TPA band at 3.7 eV as the inhomogeneously broadened $mA_g$ state in PFO [24, 44], in agreement with the EA spectra discussed above (see Table I). We emphasize that the TPA spectrum has zero strength at 3.2 eV, at the photon energy where the linear absorption spectrum has a maximum. In the semiconductor band model the VB and CB continuum bands are composed of states of both odd and even symmetries that form a band. In this case the TPA and linear absorption spectra overlap, or have very little difference, perhaps because slightly different optical dipole moments [38, 42]. In contrast, it is apparent that the TPA and linear absorption spectra *do not overlap* in PFO; in fact there is ~ 0.5 eV energy difference between their respective maxima (Fig. 4(a)). This is compelling evidence that the semiconductor band model cannot properly describe the PFO excited states. On the contrary, the energy difference between the linear and TPA spectra of ~ 0.5 eV sets the lower limit for the exciton binding energy in this polymer, in agreement with $E_b$ extracted above from the EA spectrum.

Figure 4(b) shows the polarization anisotropy in the TPA spectrum. The parallel and perpendicular TPA spectra differ by a factor of ~ 2.2; but otherwise they show the same spectral features. The polarization anisotropy found in TPA is in agreement with that of the EA spectrum. This is not surprising since both NLO spectra are related to the same $\chi^{(3)}$ thus reflecting its anisotropy, which is caused by the quasi 1D properties of the polymer chains. The two TPA spectra were fitted using the SOS model, where TPA ~ Im$\chi^{(3)}(\omega\,;\omega,-\omega,\omega)$, with the same parameters as for the fit to the EA spectrum; no extra parameter is needed. The agreement between the obtained spectrum and the model calculation is excellent. This validates the SOS model and its parameters.

*(iii) Transient photomodulation spectroscopy*

We measured the polarized pump and probe spectroscopy in a broad spectral range from 0.2 to 2.6 eV; this has never been possible before. For this extraordinary broad spectral range we used two different laser systems with PM probe spectrum in the mid ir (high repletion rate laser) and near ir/visible range (low repetition laser) [7]. The two spectra were normalized to each other in the visible range, where the PM was measured using both laser systems, either the white light continuum of the low repetition rate laser or with a doubled mid ir probe. Figure 5(a) shows the full PM spectrum of a PFO film at two delay times, t = 0 and t = 20 ps. The PM spectrum contains two relatively sharp PA bands (PA$_1$ and PA$_2$ in Fig. 5) at 0.55 and 1.65 eV, respectively, and a SE band that peaks at 2.5 eV. The three bands decay dynamics are shown in Fig. 5(b); the decay dynamics of all three bands are equal, with a lifetime of ~ 100 ps. This shows that the PM bands belong to the same photogenerated species, in contrast to earlier measurements [18]. Since stimulated emission of excitons is obtained, we then attribute this species to photogenerated singlet exciton, namely the 1B$_u$ exciton. We therefore conclude that the two PA bands related to the photogenerated excitons in PFO are transitions from $1B_u$ to the two most strongly coupled $A_g$ excitons in the singlet manifold, namely $mA_g$ and $kA_g$, respectively; no hot excitons [18] or polarons [19] are needed to interpret this spectrum. Thus the three NLO spectroscopies, namely EA, TPA and transient PM are in agreement with each other, and show that few essential states are sufficient to understand the NLO spectra in PFO.

In contrast to PPV derivatives [5, 6, 9], the PM spectrum of PFO does not show any other band that may be related to photogeneration of charge polarons. This is in agreement with the large exciton *intrachain* binding energy that we found in PFO. Also it clearly indicates that the bulky side groups of PFO do not allow for large contribution of the interchain interaction, which is largely responsible for polaron photogeneration in π-conjugated polymer films [11]. Our results are in agreement with lack of long-lived polaron photogeneration in as spun (i.e. glassy, α phase) PFO film measured by the cw PM technique [14]. Another possibility for the lack of polaron photogeneration here is the proximity of the laser excitation photon energy (~ 3.2 eV) to the absorption onset of PFO (~ 3.0 eV) [23]. From the excitation dependence of charge photogeneration in PFO using cw PM spectroscopy it was found [14] that the quantum efficiency of polaron photogeneration dramatically increases at photon energies close to $E(mA_g)$ ~ 3.7 eV; up to this photon energy there is basically very little steady state photogenerated polarons. We thus expect dramatic changes to occur in the transient PM spectrum of PFO at *higher* excitation photon energies [23], and/or when the glassy phase changes into a more ordered phase [20], and/or when a strong electric field such as in organic light emitting diodes made of PFO is capable of exciton dissociation even at low excitation photon energy [51].

## 3. Summary

We used a variety of linear and nonlinear optical spectroscopies to study the excited states and photoexcitations of the important polymer PFO with blue PL emission. The NLO spectroscopies include electroabsorption, two-photon absorption, and ultrafast photomodulation; whereas the linear spectroscopies were PL emission and absorption.

We found that the excited states of PFO are dominated by four essential states that determine the NLO spectra. These are both odd and even parity excitons: $1B_u$, $mA_g$, $nB_u$ and $kA_g$. Their energies and optical dipole moments were determined by a summation over states model that was used to fit the three obtained NLO spectra, and summarized in Fig. 6. All four states contribute to the EA spectrum; only the $A_g$ states are seen in the TPA spectrum, and the transitions from $1B_u$ to $mA_g$ and $kA_g$, respectively show up in the PM spectrum of the photogenerated excitons.

PFO cannot be described by the semiconductor band model. The best evidence for this is the comparison between the TPA and linear absorption spectra. The two spectra peak at energies that are ~ 0.5 eV apart; moreover the TPA gets a zero at the peak location of the linear absorption. This is impossible to explain by the band model. Another strong indication that PFO is excitonic in nature is the characteristic properties of the primary photoexcitations. These are excitons with two strong PA bands in the mid- and near-ir spectral range, and a SE band in the visible spectral range, rather than a typical free carrier absorption that characterizes photogenerated carriers in usual 3D semiconductors, such as Si and GaAs. Also the EA spectrum does not contain any FK oscillation at the continuum band edge, but instead shows Stark shift of the main excitons, and transfer of oscillator strength to the most strongly coupled $A_g$ states.

From the difference between $E(nB_u)$ and $E(1B_u)$ in the excited state spectrum, we estimate the lowest exciton binding energy $E_b$ ~ 0.9 eV. This is a large binding energy that indicates a strongly bound exciton. Such an excitons 'steals' most of the oscillator strength from the interband transition. Indeed there is a huge factor of ~ 100 between the excitonic transition $1A_g$ to $1B_u$ compared to the interband transition that we identify here as the $1A_g$ to $nB_u$ transition. We also obtained the polarizability of the lowest lying exciton; this is 5000 Å$^3$ that gives an exciton wave function extent of few repeat units. The seemingly contradiction between the obtained exciton wave function extent and the large binding energy may be explained by the 1D character of the excitons in PFO.

This work was supported in part by the DOE grant # FG-02 ER46109. We thank Randy Polson for the help with the measurements, and Matt Delong for the help with the sample films.

**Table I**

The best fitting parameters for the EA spectrum of PFO using the SOS model with four essential states, namely $1A_g$, $1B_u$, $mA_g$, and $nB_u$; as well as a CL distribution and phonon coupling [24]. The parameters are described in detail in ref. 24, and include the essential states' energies, $E(1B_u)$, $E(mA_g)$, and $E(nB_u)$; as well as their relative displacements, $\Delta q$ respect to each other. $h\nu$ phonon is the main vibration side band; and the CL distribution is characterized by the width $\gamma$, and asymmetry $\eta$.

| | |
|---|---|
| $E(1B_u)$ (eV) | 3.1 |
| $\Delta q_1 = q(1B_u) - q(1A_g)$ | 1 |
| $E(mA_g)$ (eV) | 3.7 |
| $\Delta q_2 = q(mA_g) - q(1B_u)$ | -0.5 |
| $E(nB_u)$ (eV) | 4.0 |
| $\Delta q_3 = q(nB_u) - q(mA_g)$ | 0.6 |
| $h\nu$ phonon (meV) | 185 |
| CL distribution width $\gamma$ (eV) | 0.2 |
| CL distribution asymmetry $\eta$ | 5 |

**Figure Captions**

Fig. 1: (Color on line) The normalized absorption and photoluminescence emission spectra of a PFO film in the glassy α phase. The polymer repeat unit is shown in the inset.

Fig. 2: (Color on line) The measured electroabsorption spectrum (crosses) of PFO, and the model fit (line) through the data points using the SOS model with fitting parameters given in Table I. The essential states $1B_u$, $mA_g$, $nB_u$ and $kA_g$ are assigned.

Fig. 3: (Color on line) The transient TPA trace (blue line) of the pump and probe pulses measured on a PFO film, compared to the pump-probe cross correlation trace measured using a NLO crystal. The relatively long-lived PA plateau is due to the PA related to the photoexcitations generated by TPA from the pump pulses.

Fig. 4: (Color on line) (a) The two-photon absorption (TPA) spectrum of PFO (circles) compared to the linear absorption spectrum (line). The essential states $1B_u$ and $mA_g$ are assigned. (b) The TPA spectrum measured with pump probe polarization either parallel (crosses) or perpendicular (circles) to each other. The line through the data points is a fit using the SOS model with parameters given in Table I.

Fig. 5: (Color on line) (a) The transient PM spectra of PFO at t = 0 (red circles), and t = 20 ps (blue circles). The PA bands $PA_1$, $PA_2$ and SE are assigned. (b) The transient decay dynamics of the main bands assigned in (a).

Fig. 6: (Color on line) The essential states energies and allowed optical transitions in PFO. The PA bands are excited state absorptions; α and TPA are linear and NLO absorption; and PL may be also replaced by SE in the pump-probe experiment.

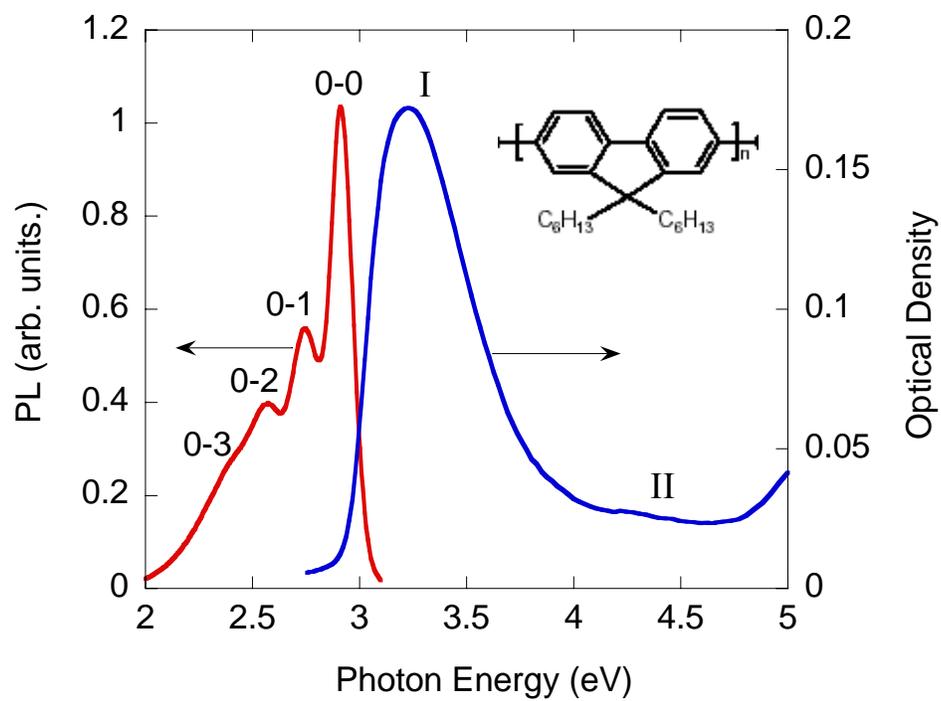

Fig. 1

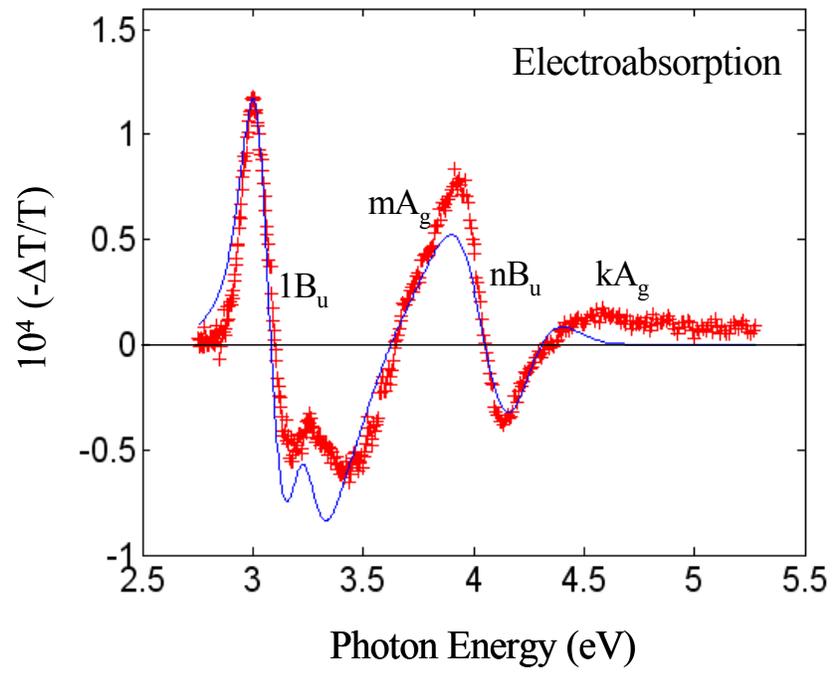

Fig. 2

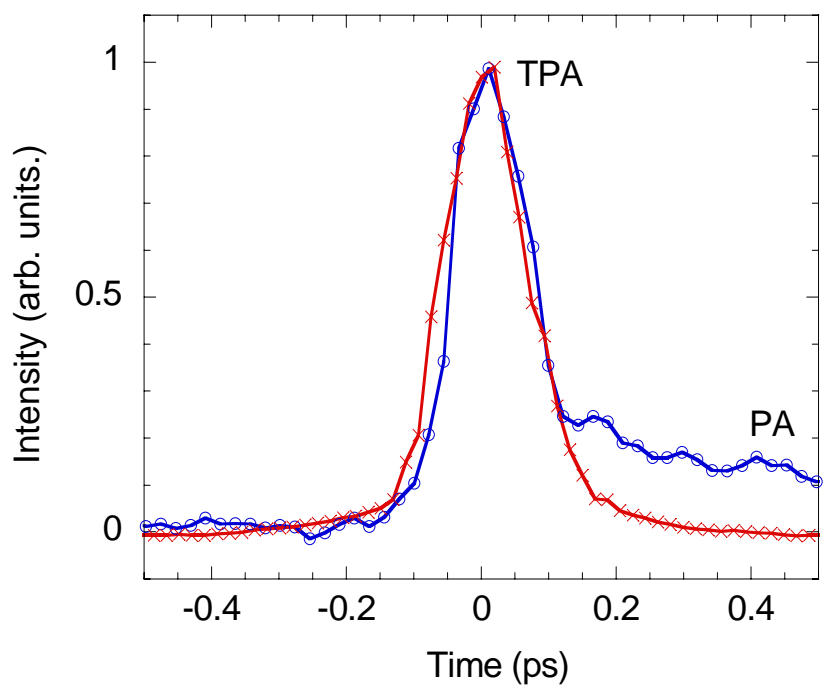

Fig. 3

(a)

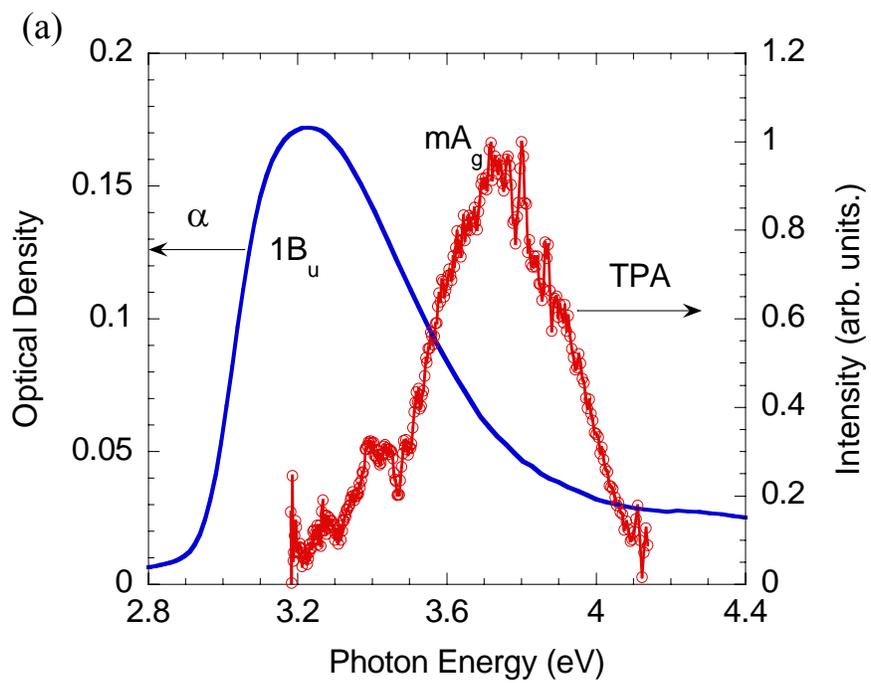

(b)

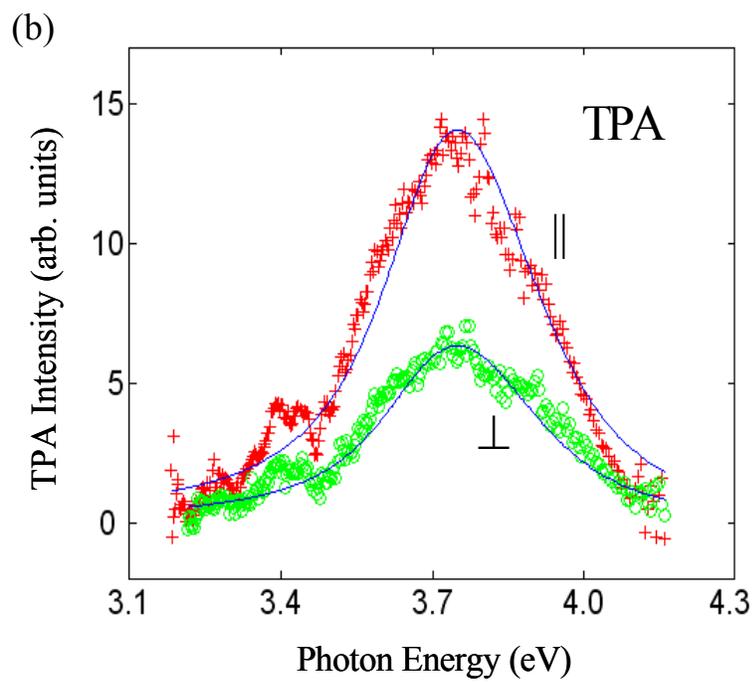

Fig. 4

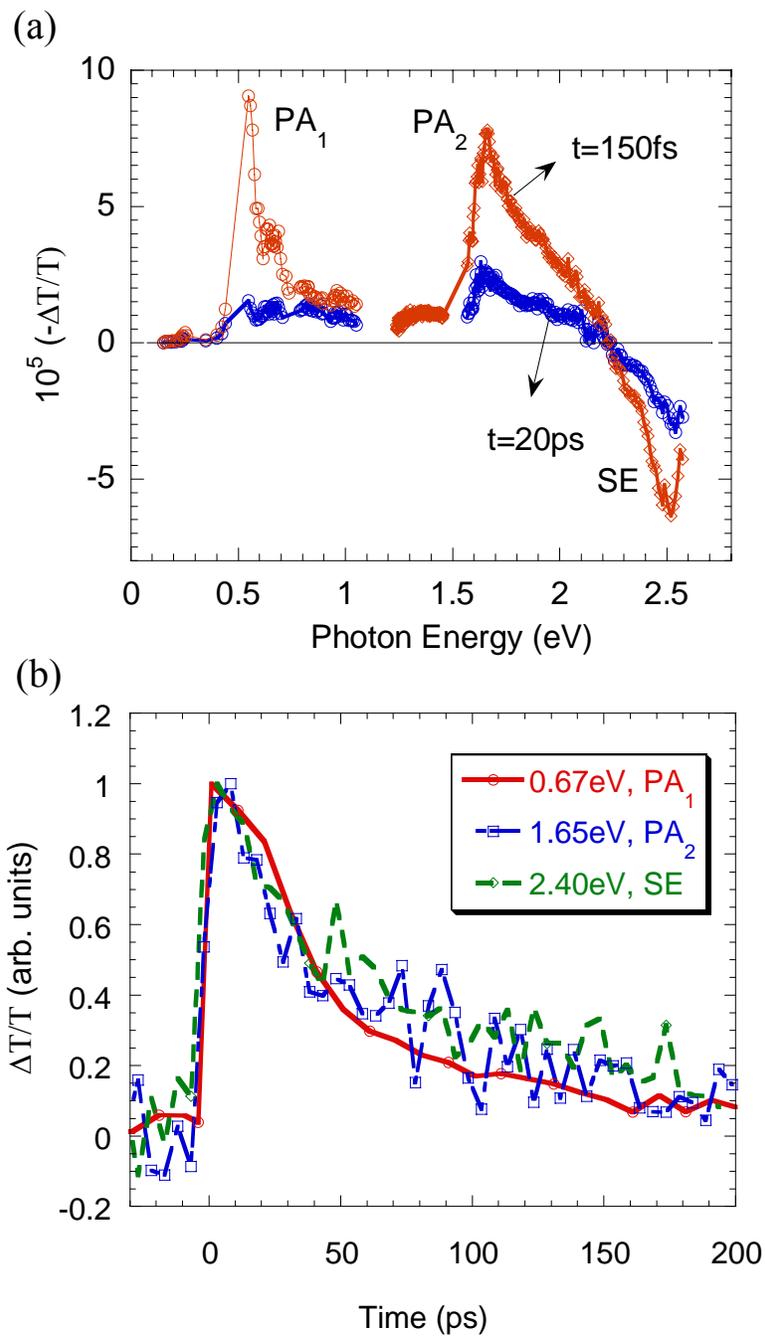

Fig. 5

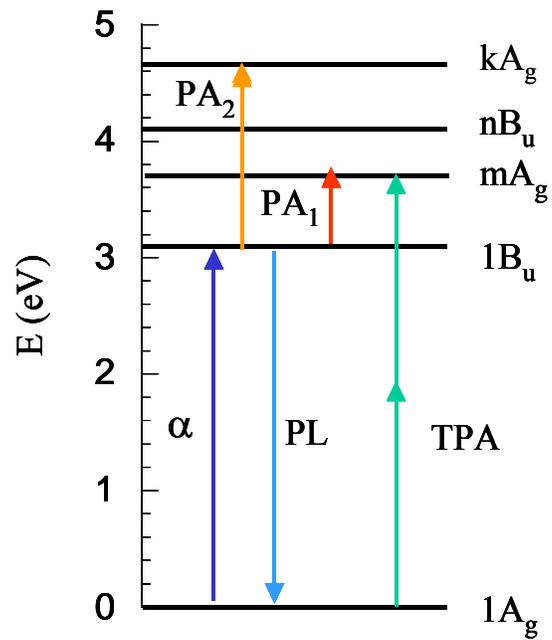

Fig. 6